\documentclass[11pt]{article}

\usepackage[utf8x]{inputenc}
\usepackage[T1]{fontenc}
\usepackage{amssymb,amsfonts,amsmath}
\usepackage{graphicx}
\usepackage{mathrsfs}
\usepackage{bm}
\usepackage{wasysym}
\usepackage{placeins}
\usepackage{lineno}
\usepackage{multirow}
\usepackage{times}
\usepackage{calc}

\usepackage{fullpage}

\title{Collapse of an ecological network in Ancient Egypt}

\author{
Justin D. Yeakel${}^{a,b,c,1,2}$, \\
Mathias M. Pires${}^{d,2}$, \\
Lars Rudolf${}^{e,f,2}$, \\
Nathaniel J. Dominy${}^{g,h}$, \\
Paul L. Koch${}^{i}$, \\
Paulo R. Guimar\~aes Jr.${}^{d}$, \\
Thilo Gross${}^{e,f}$ \\ \\
\small${}^a$Department of Ecology and Evolutionary Biology, University of California, Santa Cruz, CA 95064 \\
\small${}^b$Earth to Oceans Research Group, Simon Fraser University, Burnaby, BC, Canada V5A 1S6 \\
\small${}^c$Santa Fe Institute, Santa Fe, NM 87501 \\
\small${}^d$Departamento de Ecologia, Universidade de S\~ao Paulo, CEP 05508-090, S\~ao Paulo, SP, Brazil \\
\small${}^e$Department of Engineering and Mathematics, University of Bristol, Bristol BS8 1UB, United Kingdom \\
\small${}^f$Max Planck Institute for the Dynamics of Complex Systems, D-01187 Dresden, Germany \\
\small Departments of ${}^g$Anthropology and ${}^h$Biological Sciences, Dartmouth College, Hanover, NH 03755 \\
\small${}^i$Department of Earth and Planetary Sciences, University of California, Santa Cruz, CA 95064 \\
\small${}^1$To whom correspondence should be addressed. Email: jdyeakel@gmail.com \\
\small${}^2$J.D.Y., M.M.P., and L.R. contributed equally to this work.
}

\date{}

\begin{document}


\maketitle


\linenumbers 
\modulolinenumbers[2]

\section*{Abstract} 
The dynamics of ecosystem collapse are fundamental to determining how and why biological communities change through time, as well as the potential effects of extinctions on ecosystems. 
Here we integrate depictions of mammals from Egyptian antiquity with direct lines of paleontological and archeological evidence to infer local extinctions and community dynamics over a 6000-year span. 
The unprecedented temporal resolution of this data set enables examination of how the tandem effects of human population growth and climate change can disrupt mammalian communities. 
We show that the extinctions of mammals in Egypt were nonrandom, and that destabilizing changes in community composition coincided with abrupt aridification events and the attendant collapses of some complex societies. 
We also show that the roles of species in a community can change over time, and that persistence is predicted by measures of species sensitivity, a function of local dynamic stability. 
Our study is the first high-resolution analysis of the ecological impacts of environmental change on predator-prey networks over millennial timescales, and sheds light on the historical events that have shaped modern animal communities.

\section*{Introduction}

Modern biological communities are vestiges, with rich ecological ancestries shaped by evolutionary, climatic, and more recently anthropogenic effects.
Determining the consequences of past ecological disturbance will inform predictions of how modern communities may respond to ongoing anthropogenic or climatic pressures.
Of particular importance are extinction cascades \cite{Pace:1999dc,Eklof:2006gd}, which can lead to trophic downgrading and community collapse by altering the structure \cite{Eklof:2006gd} and relative strengths of interactions between species \cite{Estes:2011eo}.
Examining the long-term effects of extinctions on communities can only be accomplished by studying past ecosystems \cite{Yeakel:2012uc}.
The paleontological record and the remarkable historical record of species occurrences in Egypt documents a biological community changing in the face of increasing aridification and human population densities \cite{Bernhardt:2012km}.
The timing and pattern of animal extinctions in Egypt are thus well-suited to illuminate our understanding of how the structure and functioning of biotic communities are altered by changing climatic and anthropogenic impacts.


The Nile Valley north of Aswan is known for its intense heat, low rainfall, and relatively sparse vegetation.
In fact, the last 2750 km of the Nile is devoid of water-bearing tributaries and surrounded by desert with an average rainfall of 3.4 cm/yr. 
The Egyptian landscape in the Late Pleistocene/early Holocene was very different; during the African Humid Period (AHP: 14800 to 5500 yrs BP), the region had a cooler, wetter climate driven by heavy monsoonal rains \cite{Bernhardt:2012km}.
These factors contributed to a diverse assemblage of mammals that bears a strong resemblance to communities in East Africa today

Termination of the AHP was associated with increasingly weak summer monsoons \cite{Adkins:2006dw} and the disappearance of many Egyptian species, including spotted hyenas, warthogs, zebra, wildebeest, and water buffalo \cite{Churcher:1972wd,Osborn:1998wl,Linstadter:2004wa,HELLER:2012fq}, as well as the onset of dense human settlements in the region \cite{Kuper:2006dn}.
A sharp increase in aridification ca. 5000 yrs BP \cite{Kuper:2006dn,Bernhardt:2012km} attended the fall of the Uruk Kingdom in Mesopotamia \cite{Brooks:2006wv,Bernhardt:2012km}, but it might have catalyzed the rise of the Egyptian Phaoronic state \cite{Adams:1997tg,Brooks:2006wv}.
Another aridification pulse ca. $4170 \pm 50$ yrs BP \cite{Bernhardt:2012km} coincided with the Egyptian First Intermediate Period (ca. 4140 yrs BP), an interval that is distinguished by failed flooding of the Nile \cite{Butzer:2012ij} and rapid dynastic successions \cite{Trigger:1983uj}.
Other potential aridity-induced political instabilities are evident at this time, including the collapse of the Akkadian empire \cite{Cullen:2000wq} and the decline of urban centers in the Indus Valley \cite{Dixit:2014ik}.
Finally, a third aridification pulse is evident in eastern Mediterranean sediments at ca. 3000 yrs BP \cite{Bernhardt:2012km}.
This event is associated with widespread famines in Egypt and Syria \cite{Kaniewski:2008cz,Roberts:2011io} and the end of the New Kingdom in Egypt \cite{Butzer:2012ij} and the Ugarit Kingdom in Babylon \cite{Kaniewski:2008cz}.

The historical consequences of these aridification events are an enduring, and often contentious, topic of debate \cite{Demenocal:2001di,Demenocal:2004p234,Butzer:2012ib,Butzer:2012ij,Dixit:2014ik}.
At the same time, the historical ecology of Egyptian animal communities has been documented extensively \cite{Churcher:1972wd,Churcher:1993ug,Osborn:1980vi,Hendrickx:2009us}. 
These parallel efforts include descriptions of animal occurrences in paleontological, archeological, and historical records, as well as their artistic representations on tombs \cite{Davies:1943tz}, knife blades \cite{Churcher:1984tm}, and funerary palettes \cite{Spencer:1980vp}, compiled by D. J. Osborn and J. Osbornov\`a \cite{Osborn:1998wl}.
Artistic representations of mammals are identifiable at the species taxonomic level \cite{Churcher:1984tm,Osborn:1998wl}, with historical sources noting whether fauna were native or imported, or even domesticated \cite{Osborn:1998wl}. 
For example, Late Predynastic ceremonial palettes depict lions, wild dogs, and many species of ungulates including oryx, hartebeest, and giraffe (Fig. 1), none of which exist in Egypt today.
Here we combine these records of species occurrence with mathematical modeling to examine the patterns and consequences of extinctions during 6000 years of Egyptian history (Fig. 2; see SI for detailed species occurrence information). 

\section*{Patterns of Extinction}
A total of 37 large-bodied ($>4$ kg) mammalian species are documented in Late Pleistocene/early Holocene Egypt, whereas only 8 remain today \cite{Osborn:1980vi}.
Here we focus on ungulates and their potential mammalian predators, as these animals are known to form a dynamically cohesive component of many food webs \cite{Olff:2009tk,Baskerville:2010vn,Estes:2011eo,Yeakel:OfhI8s6v}, and binned their occurrences in the time periods shown in Fig. 2.
To determine whether the extinction of species in the historical record could be predicted by random removal, we conducted randomized extinction simulations ($5\times10^5$ replicates), where the number of extinctions at each time interval was conserved. 
Our analysis shows that changes in predator and prey richness -- summarized by calculating the predator-prey ratio -- are not predicted by random extinction trajectories until recent history (Fig. 3\emph{A}). 
The ratio of predators to prey increased gradually from the Late Pleistocene to the end of the New Kingdom (part of the observed increase in the predator-prey ratio after 4140 yrs BP is due to the appearance of cheetah \emph{Acinonyx jubatus}; Figs. 2 and 3\emph{A}), followed by a decline from ca. 3035 yrs BP to 100 yrs BP. 

To evaluate the effects of uncertainty in the timing of extinctions on the predator-prey ratio, we allowed the first and last occurrence of each species to vary probabilistically according to two different treatments: \emph{i}) the first/last occurrence could vary among the time-bins directly before and after the recorded event; \emph{ii}) the first/last occurrence could vary among two time-bins directly before and after the recorded event (illustrated in Fig. 2).
To determine how uncertainty influenced the predator-prey ratio, we simulated the extinction trajectories of species over time, where the occurrence of each species was drawn randomly and independently according to each extinction probability treatment (5000 replicates; Fig. 3\emph{A}).
This uncertainty introduces error in the timing of extinctions of $\pm 286$ and $\pm 580$ years (averaged across time-bins), for treatment \emph{i}) and \emph{ii}) respectively.
Importantly, we find that this added uncertainty does not alter the qualitative nature of the predator-prey ratio over time.


The loss of large-bodied herbivores, such as elephants, giraffes, native camels, oryx, and two species of kob, characterizes the earliest documented extinctions in Egypt. 
Some of these extinctions could have been caused by competitive displacement; for instance, Churcher \cite{Churcher:1972wd} suggested that wild asses (\emph{Equus asinus}), which appeared in the early-mid Holocene, might have supplanted zebras (\emph{E. grevyi} and \emph{E. quagga}, the latter formerly {\it E. burchelli}). 
Predator extinctions follow a similar pattern, with larger-bodied species disappearing earlier.
Egyptian artisans depicted two distinct lion morphotypes \cite[possibly subspecies;][]{Barnett:2014ki} before the Third Dynasty: a short-maned and a larger long-maned lion, which we treat separately. 
The long-maned lion was depicted until the end of the Second Dynasty (ca. 4645 yrs BP), and the short-maned lion until the end of the Twentieth Dynasty (ca. 3035 yrs BP; a span that excludes depictions of tame or imported lions).
Compellingly, this latter date predates accounts of diminishing lion populations in classical antiquity.
For example, lions reportedly attacked Xerxes and his consort in 2430 yrs BP, a time when lions were common in Greece (according to Herodotus, 2434-2375 yrs BP).
A little over a century later (2250 yrs BP), Aristotle reported that lions were rare \cite{Schaller:2009we}.

The most dramatic shifts in the predator-prey ratio occurred ca. 5050, 4140, 3520, 3035, and 100 yrs BP (Fig. 3\emph{A}).
Although the direction of the shift at 100 yrs BP is prone to observational error, it is coincident with population growth and industrialization in Egypt (Fig. S1). 
Three of the remaining four shifts are contemporaneous with extreme environmental and historical events: 
\emph{i}) the aridification pulse associated with beginning of the Dynastic period in Egypt \cite{Brooks:2006wv,Bernhardt:2012km} (ca. 5000 yrs BP); 
\emph{ii}) the aridification pulse associated with the collapse of the Old Kingdom in Egypt (ca. 4170$\pm$50 yrs BP); 
\emph{iii}) the aridification pulse associated with the fall of the New Kingdom in Egypt \cite{Roberts:2011io} (ca. 3000 yrs BP). 

Shifts in the predator-prey ratio reveal a long-term change in community structure: the reduction of herbivore richness beginning ca. 5000 yrs BP followed by a decline in predator richness beginning ca. 3035 yrs BP.
Although we cannot identify the causes of extinction at any single time-interval, the co-occurring changes in climate, community composition, and human societies suggest three potential mechanisms that could have resulted in the observed patterns.
First is the potential decline in herbivore richness due to human overkill followed by an indirect impact on predator richness.
Egyptian peoples shifted from mobile pastoralism after the AHP to agriculture \cite{Hassan:1985jd,Brooks:2006wv}, and subsistence hunting subsidized by agriculture \cite{Hendrickx:2009us,Doughty:2013df} may have increased overall mortality risks.
Differences in species-specific traits and hunting preferences \cite[cf.][]{Rowcliffe:2003et} would then have contributed to shape patterns of extinction.
Second, herbivore and carnivore richness may have been negatively impacted by bottom-up forcing due to climate-driven limitation in primary productivity.
Third, resource or habitat competition with humans in the Nile floodplain, driven by an increased reliance on agriculture \cite{Hassan:1985jd}, and potentially exacerbated by decreased nutrient transport from species extinctions \cite{Doughty:2013df}, might have resulted in declining herbivore richness, precipitating a cascading impact on the predator community. 

\section*{The Dynamics of Collapse}

As the composition of an ecosystem is altered, the potential dynamics of the community are bound to change \cite{Gross:2009jr}.
To understand how historical extinctions impacted the dynamics of Egyptian communities, we used predator-prey body mass ratios to calculate both the probability and strength of trophic interactions, thus reconstructing predator-prey interaction networks for each time bin \cite{Rohr:2010dy} (see Methods and SI).
We used generalized dynamical models to determine dynamic stability over time, thus requiring only basic assumptions of the functional relationships governing inter- and intraspecific interactions between and among species \cite{Gross:2009jr,Yeakel:2011p3406}. 
Across all time bins, $2\times10^5$ predator-prey networks were constructed (for parameter values and ranges, see table S1), thus accounting for potential variability in species interactions, interaction strengths, and intra- and interspecific functional responses \cite{Gross:2009jr}.
We then calculated the proportion of dynamically stable webs (PSW), the impact of a given species $i$'s presence on PSW, and the magnitude of species-specific responses to perturbations.  

Because predator-prey interactions are a function of body size, the structure of the Egyptian trophic network is relatively robust to changes in species presence/absence over time \cite[Fig. S2; cf. ref.][]{Schneider:1997tt}.
Despite the robustness of network structure, our results show that dynamic stability, measured as PSW, was highly sensitive to changes in the animal community, and reveal that extinctions in Egypt were inherently destabilizing (Fig. 3\emph{B}).
Moreover, the loss of species in the last 150 years had a disproportionately large impact on PSW (Fig. 3\emph{B}), which is a compelling indication that the effects of recent disturbances on animal communities may be more destabilizing relative to those prior to the modern era.
Stability analyses of random food webs \cite{Williams:2000wt} have generally shown that the loss of species richness increases PSW \cite{May:1972cp,Gross:2009jr}, fueling the diversity-stability debate \cite{McCann:2000wg}. 
In contrast, our analyses combining generalized modeling with a realistic interaction network structure reveal that stability decreases with species loss, and this pattern is robust against uncertainty in the timing of both historical and recent extinctions (Fig 3\emph{B}).

In the modern Egyptian predator-prey network, there are a small number of crucial species \cite{Aufderheide:2013cm} whose presence strongly and positively impacts stability, which is determined by calculating the difference in PSW ($\Delta{\rm PSW}_i$) for the system with and without each species $i$ ($2\times10^8$ replicates).
Stabilizing species include gazelles, ibex, and Barbary sheep, all of which are smaller-bodied herbivores serving as important prey resources for the remaining predators (Fig. 4\emph{A} and Fig. S3).
Some of these species (e.g. \emph{Gazella leptoceros}) are critically endangered \cite{ElAlqamy:2006jf}. 
Although the impact of species $i$'s presence on PSW is correlated with body size (Fig. S4), as we observe the community earlier in time, the presence of all species has less impact on PSW (such that $\Delta{\rm PSW}_i$ is closer to zero), suggesting that the historical community was more robust, presumably due to greater redundancy in prey species. 
Importantly, the decline in PSW essentially mirrors deviations in $\Delta{\rm PSW}_i$ away from zero, meaning that earlier communities were more stable and less impacted by species removal, while recent communities are less stable and more impacted by species removal.
Together these findings indicate an increase in vulnerability over time (Fig. S5).
We hypothesize that the vulnerability of many contemporary animal communities \cite{Estes:2011eo} may be exacerbated by recent erosion of species richness, which our data suggest eliminates the functional redundancy of lower trophic-level species. 

The primary productivity needed to support a diverse animal community is expected to have diminished as the Nile Valley became increasingly arid throughout the Holocene \cite{Butzer:2012ij}. 
Because changes in productivity can alter population-level responses to species interactions, we performed a sensitivity analysis to determine whether and to what extent changes in primary productivity influence estimates of PSW.
We address changing habitat productivity by incorporating the following assumptions:
\emph{i}) when productivity is high, the per-capita contribution of herbivores to population growth increases, such that the impact of herbivore density on growth is elevated; 
\emph{ii}) because prey are plentiful, the growth of predator populations is not limited by prey density \cite{Gross:2006wn}.  
Conversely, when primary productivity decreases (as is assumed to have occurred over the Holocene), herbivore population growth becomes nutrient-limited, such that changes in herbivore density have a smaller impact on population growth, whereas predator population growth becomes limited by herbivore density. 
This formalization allows us to explore how our results are impacted by changes in the functional relationships between population growth and its drivers due to changes in primary productivity at every time-period by instituting the following constraints:
as productivity decreases, the sensitivity of herbivore population growth to herbivore density ($\phi$ in the generalized modeling framework; see SI) goes to 0, whereas the sensitivity of predator population growth to herbivore density ($\gamma = 1 - \phi$) goes to unity; for increases in productivity, this relationship is reversed.
We find that increasing productivity is always destabilizing, which is expected in accordance with the well-known `paradox of enrichment' \cite{Rosenzweig1971}. 
Of more interest here is that lowering productivity does not have a qualitative impact on estimates of PSW (Fig. 4{\rm B}), suggesting that changes in PSW over time were not solely driven by changes in productivity itself, but were chiefly influenced by changes in community composition and species interactions.

%

\section*{Predicting Persistence}

Understanding the reciprocal feedbacks between a changing environment on the structure and functioning of ecosystems is a primary goal in modern ecological research \cite{Woodward:2010dn}.
For instance, short term environmental changes may be responsible for altering community structure in both limnetic invertebrate \cite{Schneider:1997tt} and terrestrial vertebrate food webs \cite{Lurgi:2012hf}, while shifting thermal baselines and mismatches in phenology have been observed to directly alter the composition of terrestrial communities \cite{Walther:2002fh,Sinervo:2010kk}.
Theoretical work suggests that climate warming may have a large impact on trophic chain length and top-down vs. bottom-up dynamics, while higher trophic species are predicted to be at greatest risk \cite{Brose:2012ct}. 
However to what extent the dynamical consequences of perturbed ecological communities impact species persistence is largely unknown, and this is partly due to a lack of knowledge regarding how animal assemblages and species interactions change over time \cite{BERG:2010dq}.



Although we cannot ascribe causality to any single extinction event, because the persistence of each species over time is known, we can determine whether extinction is predictable.
Perturbations are by definition disruptive, and their effects can be explored with respect to the system as a whole (PSW), or with respect to each species in the system.
In general, we would assume that species strongly reactive to external perturbations would have lower persistence, thus being prone to extinction.
Here we determine whether the sensitivities of species to external perturbations can be used to predict persistence, defined as the period of time after the Pleistocene-Holocene transition (11.7 kyr BP) of Egyptian occupation.
We define the sensitivity of a species $i$ \cite[${\rm Se}_i$; see ref.][]{Aufderheide:2013cm} by the magnitude of its response to a \emph{press perturbation}, introduced by altering the community steady state \cite{Novak:2011uz} (see SI for a formal derivation). 
Our results show that sensitivity is strongly predictive of persistence, and therefore extinction risk: species less sensitive to change are more likely to survive longer periods of time (Fig. 4\emph{C} and Fig. S6). 
Of note are two outliers for which temporal persistence is greater than predicted by ${\rm Se}_i$ (silhouettes in Fig. 4\emph{C}): Hippopotamus (\emph{Hippopotamus amphibius}), which rely primarily on river resources that are not included in the dynamic model, and wild cattle (\emph{Bos primigenius}), potentially facilitated by association with domesticates \cite{Stock:2013bw}.
Our results confirm the generally accepted expectation of higher extinction risks for larger-bodied mammalian species \cite{Cardillo:2005et}, and indicate that measures derived from local stability analysis are predictive of these risks over millennial time-scales.

The trajectory of extinctions over 6000 years of Egyptian history is a window into the influence that both climatic and anthropogenic impacts have on animal communities.
The atypically strong effects that species extinctions have had on the stability of the contemporary Egyptian predator-prey network is due to the nonrandom but steady erosion of species richness over time. 
Our results directly fuel hypotheses on whether and to what extent cascading extinctions, changes in the sensitivity to perturbations, and the consequent decline of community stability as the result of both climate change and human impact, have contributed to the collapse of modern animal communities.
\section*{Materials}
\small
We compiled data on species occurrences from paleontological, archeological, and historical information spanning the last 6000 years of Egyptian history.
All dates are yrs BP, thus `years before 1950 AD', such that we distinguish 0 yrs BP (1950 AD) from `today' (established as 2010 AD). 
We used body mass ratios between predators and prey to determine the probability that a trophic link exists between species $i$ and $j$ (${\rm Pr}(\ell_{ij} = 1)$), where ${\rm Pr}(\ell_{ij} = 1) = p/(1+p)$, given $p = {\rm exp}\{a_1 + a_2 \log ({\rm MR})+a_3 \log^2({\rm MR})\}$, and ${\rm MR}$ is the log-transformed ratio of predator to prey biomass \cite{Rohr:2010dy}.
We established this model on the Serengeti food web ($a_1=1.41,~a_2=3.73,\mbox{and~}a_3=-1.87$), from which 74\% of trophic links (both presence and absence) were predicted accurately.
We capture the dynamics of an $N$ species food web by $N$ equations of the form
$\dot{X}_i = S_i(X_i) + \eta_iF_i(X_1,\ldots,X_N) - M_i(X_i) - \sum^N_{n=1}L_{n,i}(X_1,\ldots,X_N)$,
for $i=1...N$, where $\eta_i$ is the transfer efficiency of predator growth from prey consumption; and 
$S_i$, $F_i$, $M_i$, and $L_{n,i}$ are unspecified functions that respectively describe the growth of species $i$ by primary production, the growth of species $i$ by predation, the loss of species $i$ due to natural mortality, and the loss of species $i$ due to predation by species $n$.
Local stability is computed by linearizing the nonlinear equation-system around the steady state in question. 
The result is the so-called Jacobian matrix that captures the system's response to perturbations in the vicinity of the steady state. 
For the generalized model one formally computes the linearization for all feasible steady states \cite{Gross:2009jr}. 
We thereby obtain a Jacobian matrix that captures the dynamical stability of every steady state in the whole class of models under consideration, as a function of a number of unknown, but directly interpretable ecological parameters. 
For additional details, see the online supplementary information.


\section*{Acknowledgments}
We thank S Allesina, H Aufderheide, MP Beakes, CE Chow, JA Estes, M Mangel, JW Moore, M Novak, CC Phillis, and three anonymous reviewers for helpful discussions and comments.
We also thank the Ashmolean Museum for photographic reproduction rights (Fig. 1\emph{AB}) and the British Museum for its free image service (image AN35923001 was the basis for Fig. 1\emph{C}). 
Funding was provided by the National Science Foundation Graduate Research Fellowship (NSF-GRF) to J.D.Y., and S\~ao Paulo Research Foundation (FAPESP) to M.M.P. and P.R.G. 
We also owe a debt of gratitude to C.S. Churcher and the late D.L. Osborn for the work that inspired this research.





%

\nolinenumbers




\begin{figure*}
\centering
\includegraphics[width=0.7\textwidth]{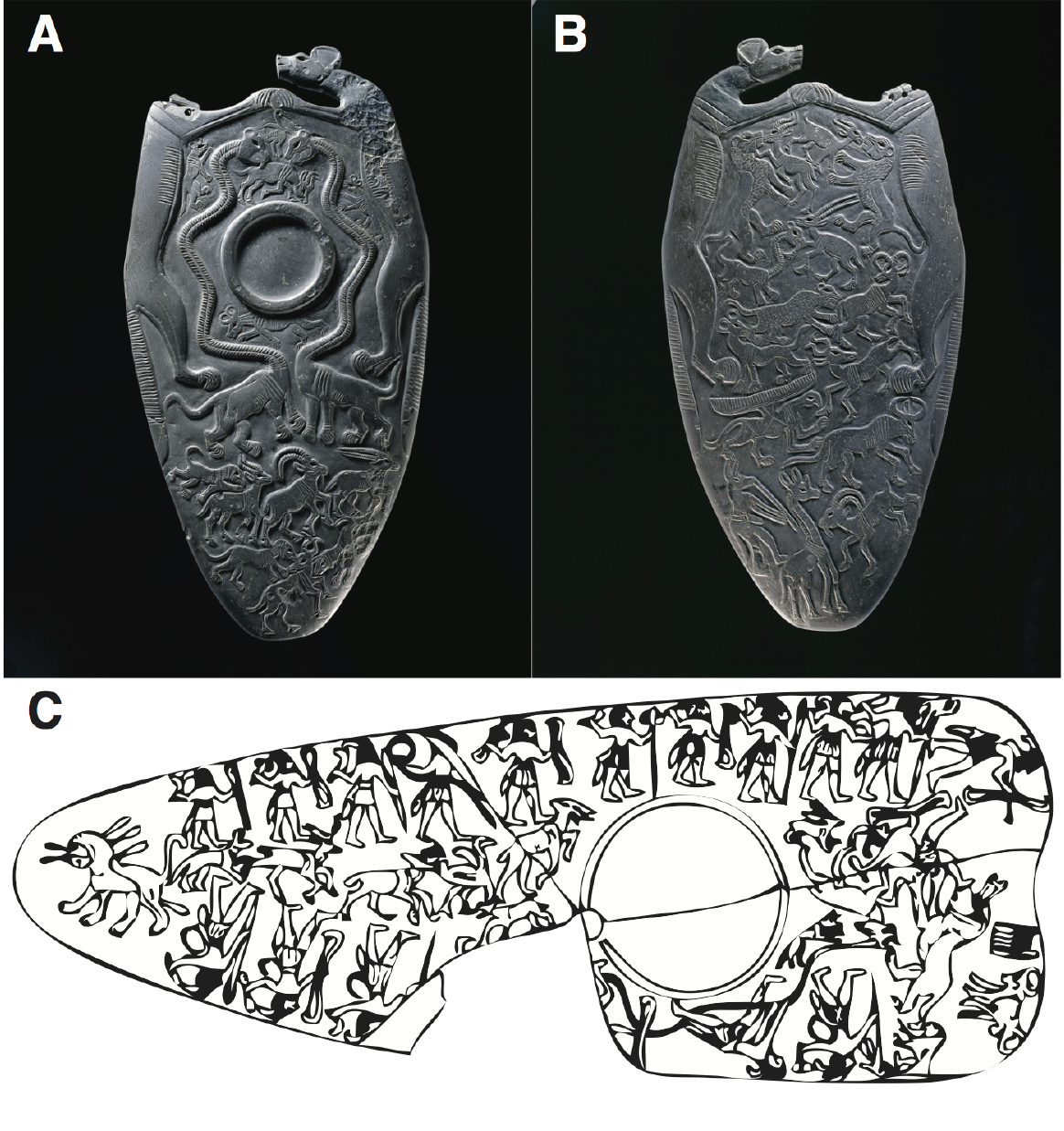}
\caption{ 
Ancient Egyptian depictions of familiar predator-prey interactions. The (\emph{A}) obverse and (\emph{B}) reverse surfaces of a siltstone ceremonial palette accessioned (no. E.3924) in the Ashmolean Museum of Art and Archaeology, University of Oxford. 
The palette (known informally as the Ashmolean or two dog palette) was recovered from the main deposit at Hierakonpolis (ca. 5150 yrs BP). 
The object is surmounted and framed by two wild dogs (\emph{Lycaon pictus}) clasping one another's paws. 
Other unambiguous species include ostrich, hartebeest, wildebeest, ibex, oryx, and giraffe. 
Some fictitious animals are also depicted, including serpent-necked panthers, or ``serpopards'', and a plausible griffin; these animals were excluded from our analysis. 
Photographs \textcopyright  Ashmolean Museum, reproduced with permission. 
(\emph{C}) Line-drawing of a mudstone ceremonial palette accessioned (no. EA20790) in the British Museum. The provenance of this Late Predynastic palette (known informally as the hunters' palette) is uncertain. 
The reliefs depict human hunters stalking and capturing lions, gazelles, hartebeest, and an ostrich with bows, spears, throwsticks and lariat.
For recent scholarship on, and interpretation of, these images, see Davis \cite{Davis:1992ve}. 
}
\label{fig:palette}
\end{figure*}

\begin{figure*}
\centering
\includegraphics[width=1\textwidth]{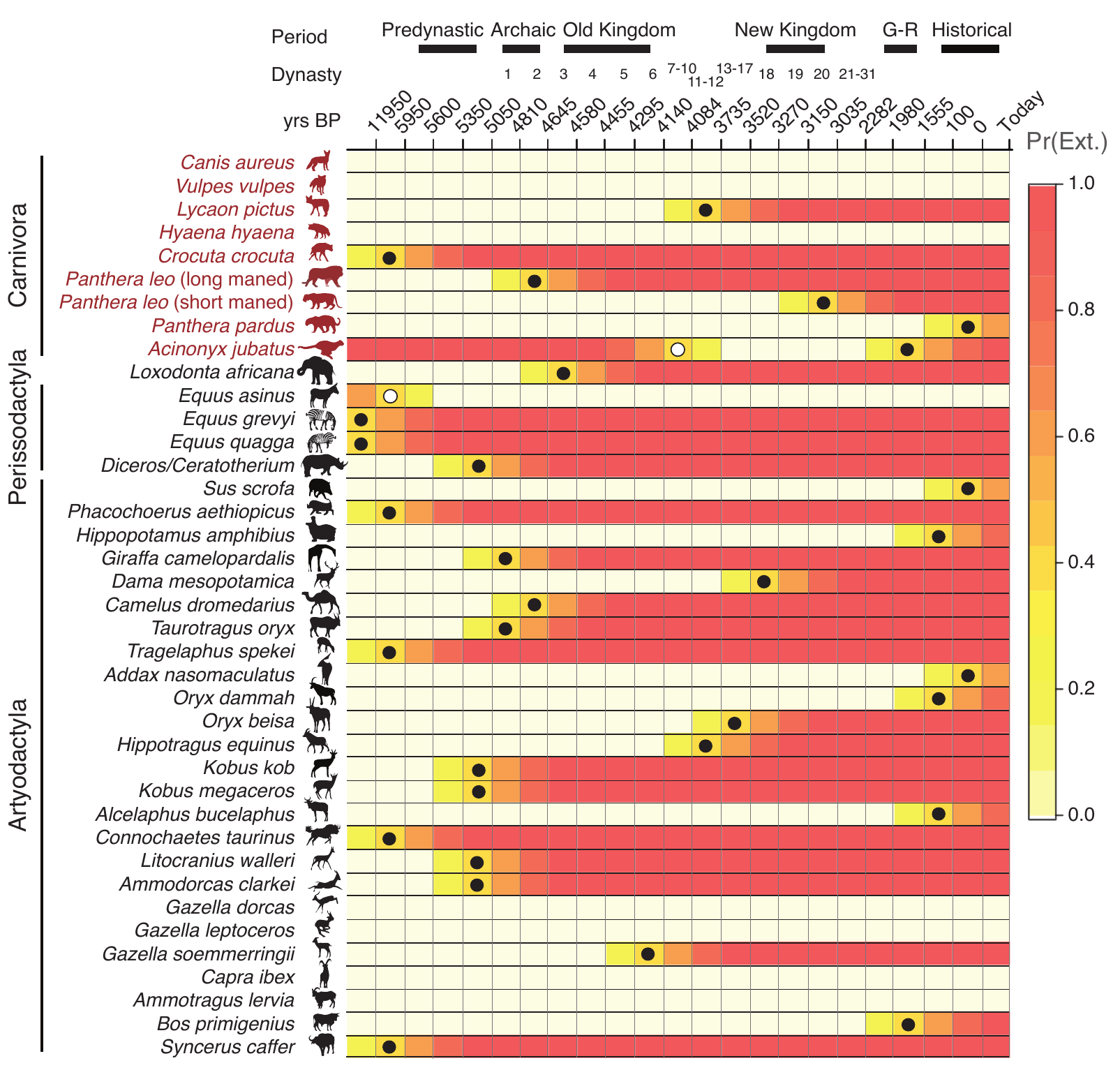}
\caption{ 
The presence/absence of large-bodied mammalian species across six millennia of Egyptian history.
All dates are yrs BP, thus `years before 1950 AD', such that we distinguish 0 yrs BP (1950 AD) from `today' (established as 2010 AD).
The first time-bin does not have a definitive starting date, generally representing the Late Pleistocene.
White circles denote the first time-interval of a recorded species occurrence if it was not initially present; black circles denote the last time-interval of a recorded species occurrence if it is not extant.
The color gradient is the probability that a given species is locally extinct for the treatment allowing first/last occupation to vary across two time-bins before and after the recorded event.
G-R = Greco-Roman.
}
\label{fig:sp}
\end{figure*}

\begin{figure*}
\centering
\includegraphics[width=0.7\textwidth]{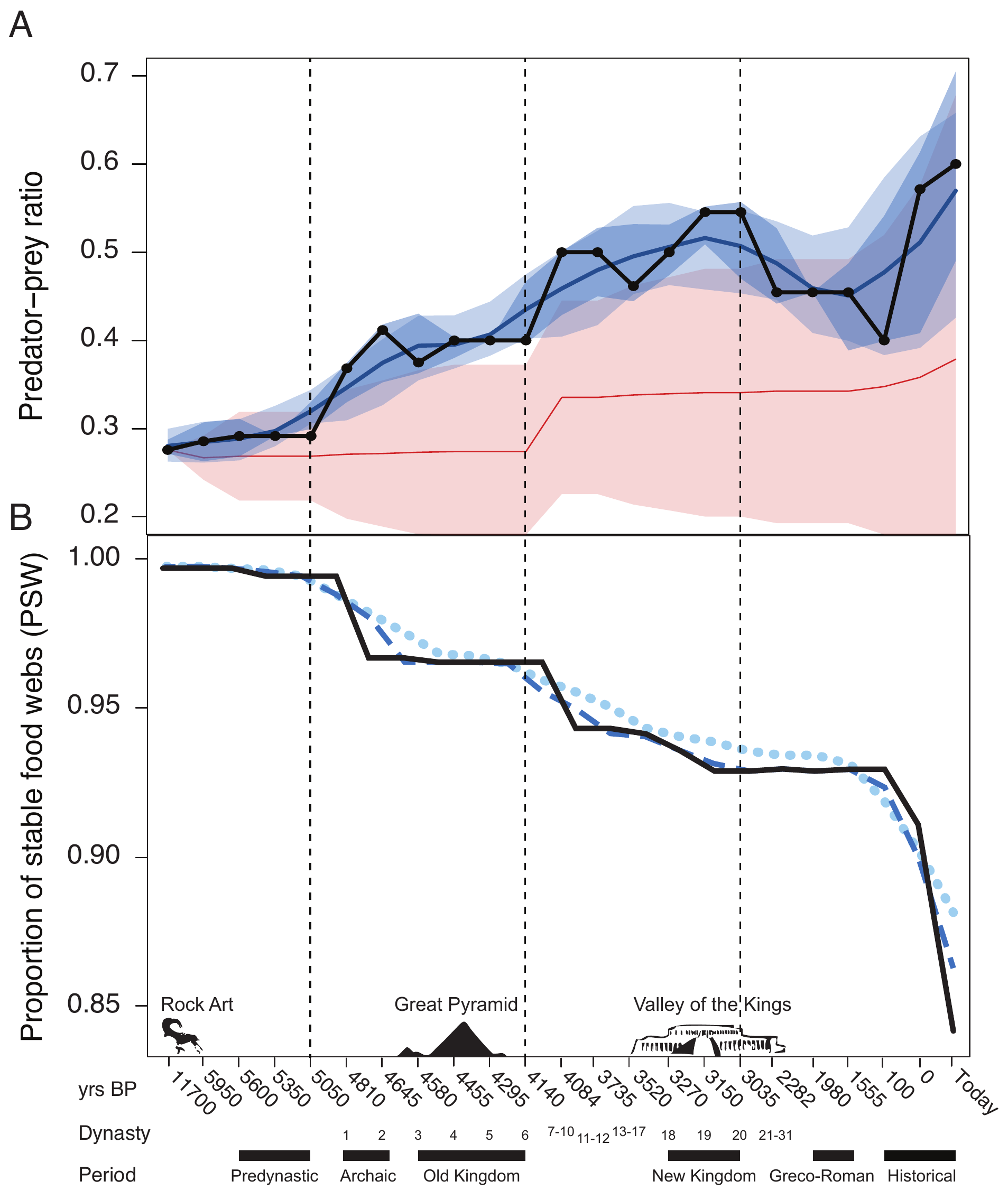}
\caption{ 
Changes in the predator-prey ratio and dynamic stability of the Egyptian trophic network over time.
(\emph{A}) The predator-prey ratio is shown (black line) against simulations where extinction is treated probabilistically, incorporating error of 286 (dark blue polygon) and 580 (lighter blue polygon) years before and after first and last occurrences (the mean is denoted by the blue line).
Random extinction trajectories are shown by the red line and polygon (mean and standard deviation of $5\times10^5$ replicates, respectively).
(\emph{B}) Proportion of stable webs ($3.2\times 10^9$ replicates) for the Egyptian community (black), and with first and last appearances treated probabilistically, incorporating error of 286 (dark blue, dashed line) and 580 (lighter blue dotted line) years before and after the first and last occurrences. 
Vertical dashed lines denote the major climatic events at ca. 5050, 4170, and 3035 yrs BP.
}
\label{fig:stable}
\end{figure*}


\begin{figure*}
\centering
\includegraphics[width=0.65\textwidth]{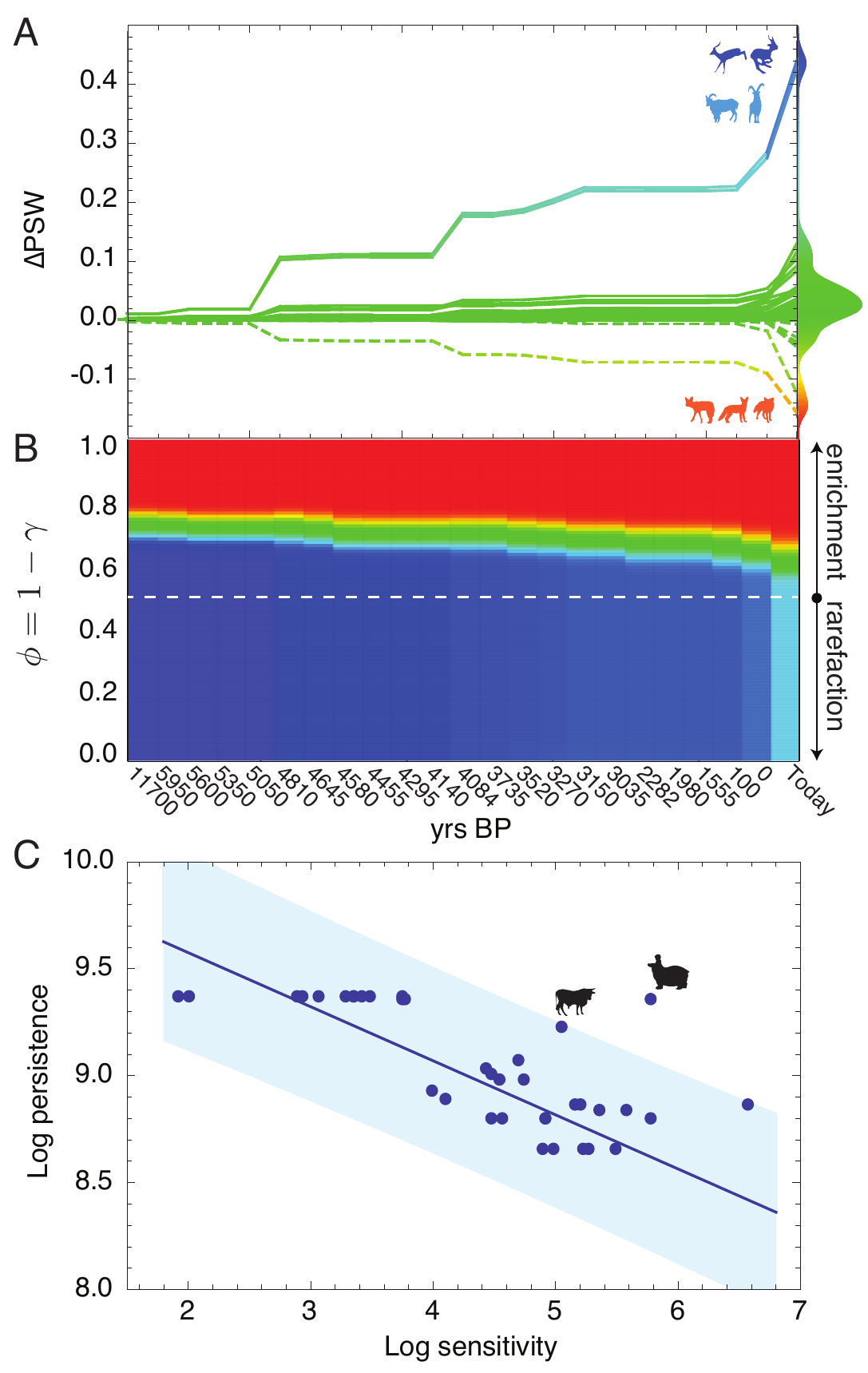}
\caption{ 
(\emph{A}) The mean change in PSW ($\Delta \rm{PSW}$) as a function of species presence over time. Predator presence destabilizes trophic network (stippled lines); prey presence stabilizes trophic networks (solid lines).
Histograms on the y-axes represent densities of $\Delta {\rm PSW}$ values for the earliest and latest time intervals, and colors scale to the y-axis.
(\emph{B}) PSW (colors scaled from red: PSW=0 to blue: PSW=1) as a function of the sensitivity of herbivore growth to changes in herbivore density ($\phi$; y-axis) over time (x-axis).
The white stippled line denotes $\phi=0.5$ used for the dynamic analysis (Table S1), whereas $\phi>0.5$ indicates nutrient enrichment (productivity increase), and $\phi<0.5$ indicates nutrient rarefaction (productivity decrease). 
(\emph{C}) Species sensitivity vs. persistence since the Pleistocene-Holocene transition (11.7 kyrs BP). 
Linear regression model: ${\rm R}^2=0.36$, $p\ll 0.005$; blue shaded region is the 75\% confidence interval. 
}
\label{fig:sens}
\end{figure*}






\end{document}